\documentclass[sigconf,nonacm]{acmart}
\usepackage{epsfig,endnotes}
\usepackage{soul}
\usepackage{url}
\usepackage{xcolor,colortbl}
\usepackage{booktabs,caption,fixltx2e}
\usepackage[flushleft]{threeparttable}
\usepackage{wrapfig}
\usepackage{float,graphicx}
\usepackage[font=small]{caption}
\usepackage{subcaption}
\usepackage{multicol}
\usepackage{multirow}
\usepackage{tikz}
\usepackage{multirow}
\usepackage{float}
\usepackage{mwe}
\usepackage{listings}
\usepackage{amsmath,mathtools}
\usepackage{array,makecell}
\usepackage{mathbbol}
\usepackage{enumitem}
\usepackage{microtype}
\usepackage{verbatim}
\usepackage{setspace}
\usepackage{bookmark}
\usepackage[ruled,linesnumbered]{algorithm2e}

\SetCommentSty{mycommfont}
\usepackage{xcolor,colortbl}
\definecolor{ashgrey}{rgb}{0.7, 0.75, 0.71}
\SetKw{KwBy}{by}
\usepackage{textcomp}
\AtBeginDocument{
  \providecommand\BibTeX{{
    \normalfont B\kern-0.5em{\scshape i\kern-0.25em b}\kern-0.8em\TeX}}}
\definecolor{Gray}{gray}{0.85}
\definecolor{lblue}{rgb}{0.7,1,1}
\definecolor{Gainsboro}{RGB}{220, 220, 220}

\newcommand{\encircle}[1]{
 \tikz[baseline=(char.base)]{\node[shape=circle,draw,inner sep=0pt] (char) {#1};}
}

\begin{document}

\title{Revisiting Logic Encryption}

\author[Rupesh~Raj~Karn, Lakshmi~Likhitha~Mankali, Zeng~Wang, Saideep~Sreekumar, Prithwish~Basu~Roy, Ozgur~Sinanoglu, Lilas~Alrahis, Johann~Knechtel]{%
Rupesh~Raj~Karn$^\ddagger$,
Lakshmi~Likhitha~Mankali$^\dagger$,
Zeng~Wang$^\dagger$,
Saideep~Sreekumar$^\ddagger$,
Prithwish~Basu~Roy$^\dagger$$^\ddagger$,
Ozgur~Sinanoglu$^\ddagger$,
Lilas~Alrahis$^\mathsection$, 
Johann~Knechtel$^\ddagger$\\[4pt]
$^\dagger$NYU Tandon School of Engineering, New York, USA\\
$^\ddagger$NYU Abu Dhabi, Abu Dhabi, UAE \\
$^\mathsection$Computer and Information Engineering, Khalifa University, Abu Dhabi, UAE \\[3pt]
\text{\{rupesh.k, likhitha.mankali, zw3464, sds710, pb2718\}@nyu.edu}
\text{ozgursin@nyu.edu, lilas.malrahis@ku.ac.ae, johann@nyu.edu}
}

\begin{abstract}
Modern circuits face various threats like reverse engineering, theft of intellectual property (IP), side-channel attacks, etc.
Here, we present a novel approach for IP protection based on logic encryption (LE).
Unlike established schemes for logic locking, our work
obfuscates the circuit's structure and functionality by encoding and encrypting the logic itself.
We devise an end-to-end method
for practical LE implementation
based on standard cryptographic algorithms, key-bit randomization, simple circuit design techniques, and system-level synthesis operations,
all in a correct-by-construction manner.
Our extensive analysis demonstrates the remarkable efficacy of our scheme, outperforming prior art against a range of oracle-less attacks covering crucial
threat vectors, all with lower design overheads.
We provide
a full open-source release.
\end{abstract}


\maketitle

\section{Introduction}
\label{sec:introduction}

Protecting integrated circuits (ICs) against intellectual property (IP) piracy and reverse engineering (RE), amongst other threats, has become increasingly vital in today's distributed landscape for design and
manufacturing~\cite{chakraborty2019keynote,akter2023survey,tehranipoor2023hardware,xue2020ten,tan2020benchmarkingfrontierhardwaresecurity}.

The term \textit{logic encryption (LE)} was originally coined for various techniques that embed additional key-gate (KG) structures into IC designs~\cite{epic,zhou2019resolving,gandhi2023logic};
correct operation of the IC requires the correct key-bit for these structures.
While those early works established a solid foundation for protection of modern IC designs,
      unlike the wording suggests,
      they did not pursue actual encryption of the logic.

\begin{figure}[tb]
\centering
		\includegraphics[width=.78\columnwidth]{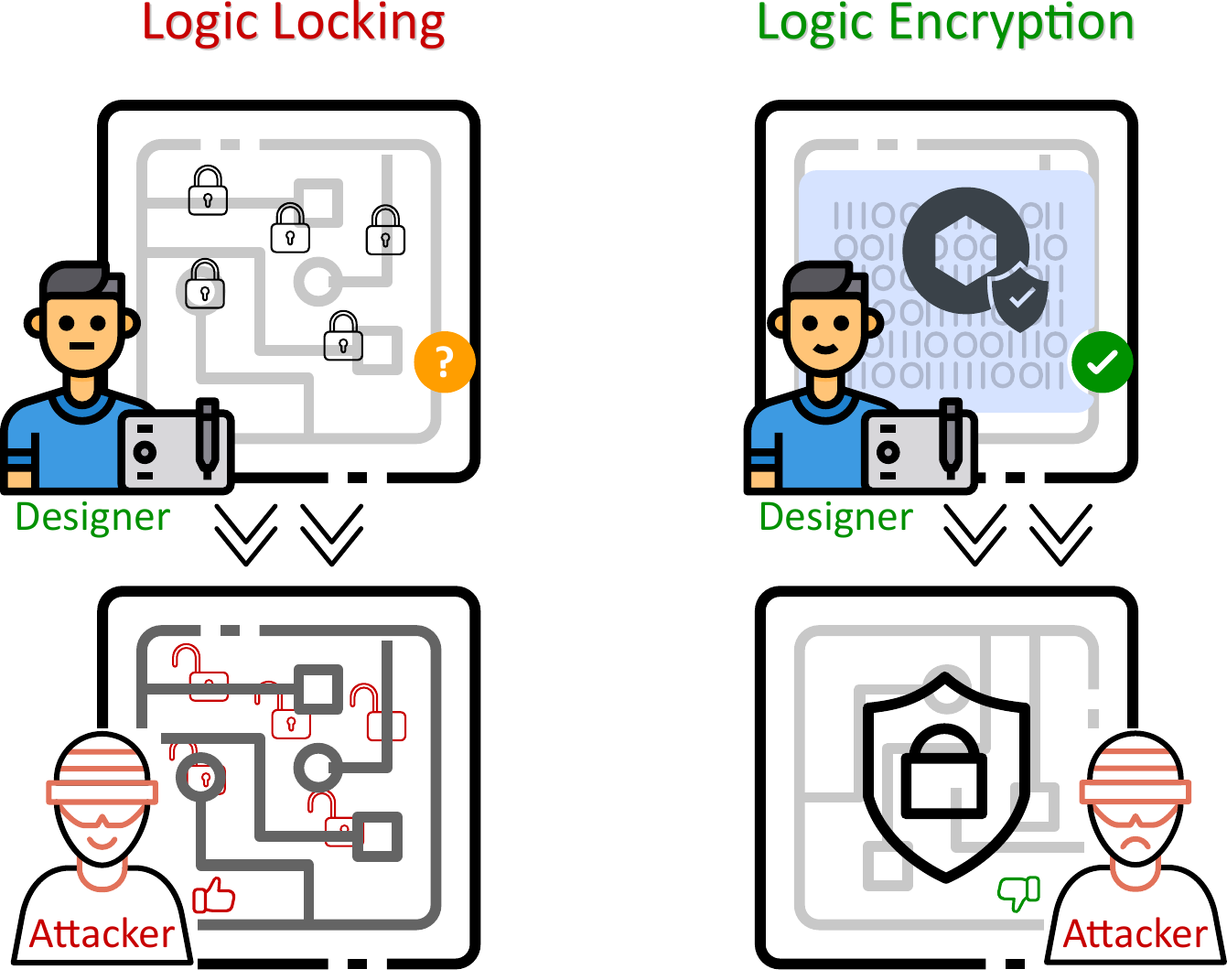} 
		\caption{Our approach for revisiting LE is fundamentally different from prior art of LL. Instead of integrating KG structures into the netlist as an afterthought,
			our work encrypts the netlist itself, utilizing cryptographic algorithms and simple circuit design techniques, all to fully obfuscate the design IP with competitive overheads.}
		\label{fig:high_level}
\end{figure}

Nowadays, this approach is more commonly known as \textit{logic locking (LL)}.
By integrating  KG structures within the netlist more as an afterthought, LL can fall short in truly protecting the design IP.
In fact, numerous attacks have shown, time and again, significant vulnerabilities across LL schemes,
   e.g.,~\cite{tan2020benchmarkingfrontierhardwaresecurity,sirone2020functional,ahmed2024seemless,raj2023deepattack,aksoy2024kratt,moosa2023scaling,resynth,omla,muxlink,scope, chakraborty2018sail, tehranipoor2024advances,vojvoda2024experimental, wang2025optilock, wang2023autolock, rajendran2012security, alrahis2021gnnunlock}.
Our work signifies a paradigm shift, back to the roots of the wording.
For the first time,
    we advocate for encryption of the logic itself, aiming for full-scale obfuscation of the design IP.
Our approach to LE involves encoding the original circuit into binary plaintexts, which are encrypted using established cryptographic algorithms. The resulting ciphertexts are
decoded back into an encrypted circuit. The latter is integrated with some decryption-like circuitry into the final protected design.
Full details for our method are provided in Sec.~\ref{sec:method} and our release~\cite{release}.
We illustrate the high-level difference between established LL techniques and our approach to LE in Fig.~\ref{fig:high_level}.

In short, our paper provides several contributions:
\begin{enumerate}

\item \textit{Logic Encryption}:
Revisiting the foundation of logic encryption/locking, our approach emphasizes on encrypting the logic itself.
Despite the long history of the field, this is a first-of-its-kind work.
The motivation is to fully obfuscate the design IP against advanced attacks.

 \item \textit{End-to-End, Commercial-Grade, Open-Source Method}: We devise and implement a comprehensive end-to-end method for realizing LE.
 Our design analysis, and our method as a whole,
 utilize commercial-grade CAD tools, rendering it robust and relevant for real-world insights.
 We provide a full open-source release at \cite{release}.
 
 \item \textit{Thorough Security and Design Evaluation}: We conduct an extensive evaluation of our approach for both security and power, performance, and area (PPA).
We also compare ours to state-of-the-art (SOTA) for LL~\cite{trll,dmux,lut}.
 Our security analysis is
 based on open-source attack frameworks and covers
 critical threat vectors for the oracle-less setting:
 \begin{enumerate}
 \item machine learning (ML)-based prediction of KG structures~\cite{omla},
 \item ML-based prediction of KG interconnects~\cite{muxlink},
 \item resynthesis-based exploitation of information leakage by KG structures and interconnects~\cite{resynth,scope},
 \item and ML-based RE~\cite{GNNRE}.
 \end{enumerate}
\end{enumerate}

\section{Preliminaries}
\label{sec:prelim}

We begin with a concise overview of several key attacks—\textbf{OMLA~\cite{omla}}, \textbf{TRLL~\cite{trll}}, \textbf{gDMUX~\cite{dmux}}, \textbf{LUT-L~\cite{baumgarten10}}, MuxLink~\cite{muxlink}, GNN-RE~\cite{GNNRE} and the resynthesis attack~\cite{resynth}. This foundation will facilitate the reader's comprehension before we present the implementation of these attacks on our logic encryption scheme in Sections~\ref{subsec:security_analysis_setting} to \ref{subsec:security_analysis_gnnre}.

A plethora of LL schemes exist, e.g., \cite{
tan2020benchmarkingfrontierhardwaresecurity,
saxena2024efficient,dupuis2024logic,trll,lut,dmux,wang23,interlock,rajendran2013fault,8341984}, but most have been challenged over time \cite{tan2020benchmarkingfrontierhardwaresecurity,sirone2020functional,ahmed2024seemless,raj2023deepattack,aksoy2024kratt,moosa2023scaling,resynth,omla,muxlink,scope, chakraborty2018sail,
tehranipoor2024advances,vojvoda2024experimental, rajendran2012security, alrahis2021gnnunlock, el2025graph}.
While early attacks followed the oracle-guided threat model, e.g., \cite{yang2019stripped,subramanyan2015evaluating},
the recent rise of oracle-less attacks
now dominates \cite{omla,resynth,muxlink,scope,karmakar2024noball,aghamohammadi2024lipstick,insight,resaa}
Thus, in this work, we also focus on the oracle-less threat model.

    In traditional LL with XOR/XNOR KGs,
	    information leakage arises from the KG type directly mapping to the correct key-bit: 0 for XOR and 1 for XNOR~\cite{epic}.
Logic synthesis is employed after KG insertion in an effort to disrupt this mapping.
Still, ML-based attacks like OMLA~\cite{omla} and resynthesis attack~\cite{resynth} show that such synthesis efforts are deterministic, with
predictable circuit modifications depending on the key-bits, and their effects being localized.

\textbf{Advanced LL Schemes:}
Based on these insights, researchers strive for LL schemes
	that do not rely on synthesis for resilience.
	Representative schemes are outlined next.

Truly random LL (TRLL)~\cite{trll} performs its own bubble pushing and inverter absorption to
insert XOR/XNOR KGs, aiming for
a truly randomized correlation between key-bits
and KG types.

LL may utilize multiplexers (MUXes) for their inherent structural ambiguity~\cite{wang23,dmux}.
For example, a 2-to-1 MUX acting as KG takes two data inputs, a true wire and a false wire,
and a key-input drives the select line.
As the true wire can be randomly connected either to the first or second data input of the MUX, the correct key-bit is truly random 0 or 1, with no specific key-bit associated with the MUX KG as such.
The scheme gDMUX~\cite{dmux} goes further by strategically constructing pairs of MUX KGs such that the circuit remains fully connected regardless of the key-bit applied.
LL may also utilize key-controlled look-up tables (LUT)
	to redact gates or whole blocks from the circuit~\cite{baumgarten10};

we refer to this scheme as LUT-L in the remainder.

Another kind of redaction is proposed in~\cite{saha20}: their idea is to embed circuits into cryptographic hardware primitives, thereby locking the circuits' IP, but not encrypting it.\footnote{
Their approach is orthogonal to ours for multiple reasons. (1)~They embed circuits into cryptographic
hardware, whereas we utilize cryptographic algorithms for actual circuit/logic
encryption.  (2)~Their applicability is severely limited: less than 10 gates are
redacted across all experiments in~\cite{saha20}, yet with significant PPA
overheads.  (3)~Theirs was proposed exclusively for the oracle-guided threat
model. In that context, another attack~\cite{roy24} questioned their security against power
side-channel attacks.}

\textbf{Advanced Attacks:}
Despite their promises, these LL schemes remain vulnerable.
	Representative attacks are outlined next.

SCOPE~\cite{scope} explores both possible key-bit values, 0 vs.\ 1, for each KG structure and, for each possible combination of key-bit values across all KG structures, runs simple resynthesis efforts.
It then infers any correlation between the key-bit values and the resulting PPA features, e.g., smaller area due to
removal of redundant logic induced by incorrect key-bit values.

The resynthesis attack~\cite{resynth} advances this approach, by first running synthesis for various settings, including challenging timing constraints on KG structures, and subsequently providing SCOPE with many more structural variations to explore.
Recently, this approach was successfully extended to compound LL schemes,\footnote{%
The LL schemes described above aim for resilience in the oracle-less threat model, but not the oracle-guided threat model, whereas other schemes like~\cite{saha20,interlock}, often referred to as ``provably secure LL'', aim for the reverse.
Compound locking integrates schemes from both worlds, aiming for more holistic defenses
\cite{tehranipoor2024advances,chakraborty2019keynote},
}
realizing a divide-and-conquer strategy, as in first separating the different LL schemes and then tackling them separately~\cite{resaa}.

For TRLL, while KG structures are better obfuscated, more intricate
circuit characteristics, such as the number of inverters and the overall distribution of gate types, still allowed advanced ML-based attacks
to succeed~\cite{insight}.

MuxLink~\cite{muxlink} is tailored for
MUX-based locking; it
considers the challenge of deciphering the correct key-bit as a link-prediction problem.
MuxLink achieves accuracy as high as 100\% on gDMUX and other SOTA schemes.
\textbf{Reverse Engineering:}
Another important aspect, which is often overlooked, is whether LL can also prevent RE.
Note that RE is less stringent than key-recovery attacks; RE aims to understand the overall functionality of the design, not necessarily all key-bits and full circuit details~\cite{GNNRE, bucher2022appgnn}.

GNN-RE~\cite{GNNRE} provides a modern approach to RE; it
trains a graph neural network (GNN) on a circuit dataset with structural variations, e.g., comprising different adder implementations,
to identify the functionality of unseen IP.
Notably, GNN-RE is able to identify IP even with LL in place~\cite{GNNRE}.

\section{Method for Logic Encryption}
\label{sec:method}

\begin{figure*}[tb]
\includegraphics[width=0.80\textwidth]{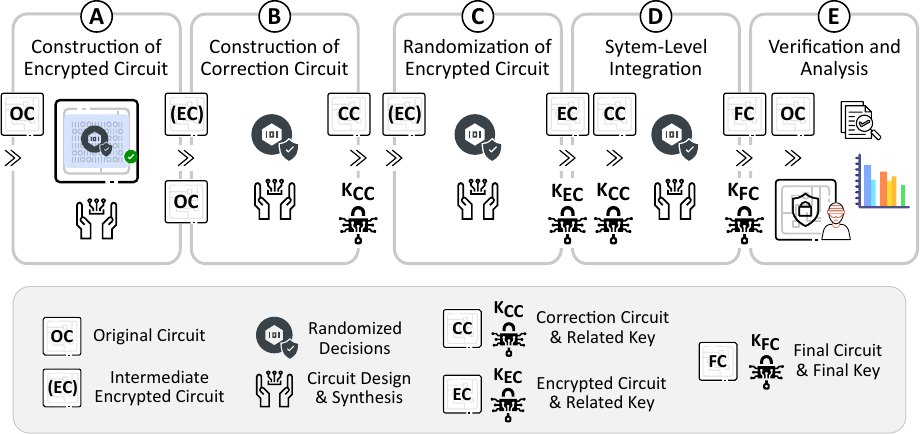} 
		\caption{End-to-end method for LE, separated into key stages.
}
		\label{fig:method}
\end{figure*}

The fully automated, end-to-end method for LE is outlined in Fig.~\ref{fig:method}.
Next, we provide details for each stage.
Note that implementation examples are provided in our release at \cite{release}.
Furthermore note that, while we utilize synthesis throughout all stages, this is only for implementation efforts---the resilience of our scheme does not depend on synthesis obfuscating KG structures. We also demonstrate
this in our security analysis in Sec.~\ref{sec:experiments}.

\subsection{Construction of Encrypted Circuit}
\label{sec:method:EC}

The first stage, labelled \encircle{A} in Fig.~\ref{fig:method}, is the construction of the encrypted circuit (EC).
The EC is a core component for LE, along the correction circuit (Sec.~\ref{sec:method:CC}).

\textbf{1) Devise Coding Scheme:}
First, the original circuit (OC) is
prepared for encryption. We must encode all gates such that subsequent decoding (after encryption) is complete and deterministic. Thus, we require a coding
scheme where the set of code-words fully covers the set of gate types. For simplicity, we also require that all gate types have the same number of inputs/outputs, to maintain
the original interconnects as such.

In this work, for simplicity but without loss of generality (w/o.l.o.g.), we utilize the set of universal NAND and NOR gates.
Importantly, this simple coding scheme does not undermine the subsequent cryptography-enabled obfuscation of the OC.
The actual coding scheme, i.e.,
the assignment of code-words 0 and 1 to NAND and NOR, or vice versa, is randomly decided and memorized.

\textbf{2) Resynthesize OC and Encoding of Gates:}
Once the coding scheme is finalized, the OC is resynthesized using only the gate types supported in the scheme.
Then, each gate in the resynthesized OC is encoded. The order in which all gates are selected during encoding is randomly decided and memorized. The result of this step is a binary plaintext, of
length $g$ bits (in this work) for $g$ gates in the resynthesized OC.

\textbf{3) Encryption of Encoded Gates:}
The plaintext, i.e., the encoded representation of the resynthesized OC, is encrypted.
W/o.l.o.g., we utilize AES-128 with random keys.\footnote{%
We do not memorize these keys, as we do not decrypt the EC later on; see Sec.~\ref{sec:method:CC}.}
The plaintext of $g$ bits in total is split into 128-bit messages, and the last message is padded with random bits as needed.
The result of this step is a binary ciphertext,
    representing the EC.

{This step ensures that attackers cannot decipher the true functionality of the OC from the EC without the encryption key.}
Based on the well-known cryptographic principles of confusion and diffusion, any correlations between the logic of the OC and EC are truly obfuscated.
In short, \ul{we utilize cryptography for truly randomized obfuscation of the entire OC and all its IP, which is a major difference over prior art in LL.}

\textbf{4) Decoding of Gates into EC:}
 Utilizing the coding scheme, the ciphertext, and the memorized order of gates selected during encoding, all gates are decoded. That is, for every gate $g_i$, the corresponding bit $c_i$ of the ciphertext is decoded into
 either a NAND or NOR gate. The result of this step is the basic EC.

\textbf{5) Resynthesize EC:}
To maintain PPA efficiency, the basic EC is resynthesized using the full library.
	Importantly, this step does not undermine the functional-centric circuit transformations realized by encryption,
	as resynthesis guarantees functional equivalence by construction.
The result of this step is the intermediate EC, as required for the next stage (Sec.~\ref{sec:method:CC}), whereas the final EC is obtained only later (Sec.~\ref{sec:method:rand_EC}).

\subsection{Construction of Correction Circuit}
\label{sec:method:CC}

Next, labelled \encircle{B} in Fig.~\ref{fig:method}, we construct the correction circuit (CC).
The CC ensures that the final circuit (FC) can realize the intended functionality despite the encrypted logic.
Importantly, {the CC is \textit{not} constructed by decryption of the EC;}
rather, the CC is constructed such that it operates on the primary outputs (POs) of the EC for correction of the functional behaviour.

\textbf{1) Devise CC Logic, Including Random Keys:}
 Let \( PO_{OC} \) denote a PO of the OC and \( PO_{EC} \) the corresponding PO of the EC. Then, the logic for the corresponding PO of the CC is obtained as

\(
PO_{CC} = PO_{OC} \oplus PO_{EC}'
\)

where \( \oplus \) denotes the XOR operation and
\( PO_{EC}' \)
is obtained from
\( PO_{EC} \) in a randomized manner, either as-is (buffered) or inverted.
Importantly, these random choices are memorized for all POs, as so-called intermediate key-bits $K_{CC}$, whose value is 0 for buffering / 1 for inverting the logic, respectively.

\textbf{2) Synthesize CC:}
For all POs, the above logic is compiled into a top-level \textit{Verilog} module, also called wrapper. The wrapper is synthesized with the OC and the intermediate EC as instantiated modules.
The result of this step is the final CC.
\subsection{Randomization of Encrypted Circuit}
\label{sec:method:rand_EC}

The third stage, labelled as \encircle{C} in Fig.~\ref{fig:method}, is to randomize the EC.
The motivation here is as above, i.e., to obtain some further
key-bits
essential for correct operation of the FC.

\textbf{1) Devise EC Logic, Including Random Keys:}
Let \( PO_{EC} \) denote a PO of the intermediate EC. Then, its reassignment can be expressed as
\(
PO_{EC,\text{new}} = PO'_{EC}
\)
where \( PO'_{EC} \) is obtained from
\( PO_{EC} \) in a randomized manner, either as-is (buffered) or inverted.
As before, these random choices are memorized, as intermediate key-bits $K_{EC}$ of value 0 for buffering / 1 for inverting the logic, respectively.

\textbf{2) Synthesize Final EC:}
For all POs, the above logic is compiled into another wrapper, which is synthesized with the intermediate EC as module.
The result of this step is the final EC.

\subsection{System Integration}
\label{sec:method:integration}

\begin{figure}[tb]
\centering
		\includegraphics[width=.75\columnwidth]{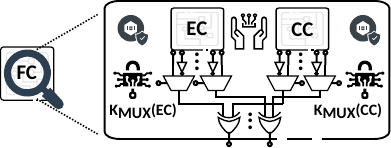} 
		\caption{Obfuscation of system-level interconnects between the encrypted circuit and the correction circuit.}
		\label{fig:LE_MUX}
\end{figure}

The fourth stage, labelled as \encircle{D} in Fig.~\ref{fig:method}, is the secure system-level integration of the EC and the CC into the FC.
Toward that end, we utilize MUX-based KG structures that introduce yet another layer of obfuscation, namely for the correct connections between the EC and the CC.
These KG structures are shown in Fig.~\ref{fig:LE_MUX} and explained in detail next.
Importantly, \ul{our use of KG structures here differs from prior art as they are not introduced throughout the netlist as afterthought, but rather as key components at the system level.}

\textbf{1) Devise FC Logic, Including Random Keys:}
We feed any pair of corresponding POs from the EC and the CC to individual XOR gates, which in turn drive the FC's POs.
This enables restoration of the intended functionality as follows:
\begin{align}
PO_{FC} & = PO_{EC} \oplus PO_{CC} \\
	& = PO_{EC} \oplus \left( PO_{OC} \oplus PO_{EC} \right) = PO_{OC}
\end{align}

Note that we are not feeding the EC/CC POs directly to XOR gates, but first pass them to the MUX KG structures.
More specifically, for both EC and CC POs, we randomly and independently assign the respective buffered/inverted PO logic to the two data inputs of these KG structures.
These randomized decisions are memorized as intermediate key-bits $K_{MUX}$, which are of value 0 for the buffered PO logic connected to the first data input and the inverted PO logic to the second data input, and of value 1 for the
vice-versa assignment.
Importantly, the correct key-bit values could be either 0 or 1 here,
	{both depending on these randomized decisions for the MUX KG
construction as well as the previously made randomized decisions to buffer or invert the EC/CC POs during their construction.} In simpler words, the different intermediate key-bits are entangled by construction.

\textbf{2) Derive Final Keys:}
Altogether, we have three sets of intermediate key-bits, $K_{EC}$, $K_{CC}$, and $K_{MUX}$.
	Note that, while $K_{EC}$ and $K_{CC}$ are of size $|PO|$, $K_{MUX}$ and $K_{FC}$ are of size $2\times|PO|$, with $|PO|$ being the same across OC, EC,
CC, and FC.
The final key-bits, $K_{FC}$, are independently derived for each PO as follows, considering the entanglement of the intermediate key-bits:
\begin{align}
K_{FC}(PO_{EC}) &= K_{MUX}(PO_{EC}) \oplus K_{EC}(PO) \\
K_{FC}(PO_{CC}) &= K_{MUX}(PO_{CC}) \oplus K_{CC}(PO)
\end{align}

\textbf{3) Synthesize FC:}
For all POs, the above logic for EC and CC integration with MUX KG structures and XOR assignments is compiled into yet another wrapper.
The wrapper is synthesized with the final EC and the final CC as instantiated modules.
The result of this step is the final circuit, FC.

\subsection{Verification and Analysis}

The final stage, labelled as \encircle{E} in Fig.~\ref{fig:method}, serves formal verification and analysis.
Formal verification is motivated by the many steps incorporating some randomized transformations in our scheme, ranging from encryption (with the coding scheme, gate selection during
encoding, and the encryption key all being randomized), over construction of the EC and CC,
to the system-level integration of the FC, all with their own 
intermediate key-bits.
Still, {all steps are correct-by-construction}
-- verification is only employed as ``sanity check'' for our fully automated end-to-end flow.
In fact, all LE runs pass verification;
		proofs are provided in our release \cite{release}.

\subsection{Implementation}

\textbf{Coding:}
The encryption stage
and all processes for compiling the different wrappers are implemented in \textit{Python}.
Synthesis processes are driven by \textit{tcl} scripts, with synthesis itself conducted by Synopsys DC.
Data management, including arrangement of the various circuit netlists across the different stages,

	  is implemented by \textit{bash} scripts.
Importantly, the whole method is fully automated.

\textbf{Tooling:}
We utilize Cadence Conformal for verification, and
Synopsys DC for synthesis and design analysis.
Further details, in particular for the security analysis, are provided in Sec.~\ref{sec:experiments}.

\textbf{Runtimes:}
Running the method end-to-end requires only tens of minutes across all benchmarks.
Since all key steps for circuit design are underpinned by commercial-grade synthesis,
runtimes scale graciously for larger benchmarks.

\textbf{Release:}
We provide a full open-source release of our method at \cite{release}.
This includes all scripts/codes, as well as end-to-end examples for LE implementations on all benchmarks.\footnote{%
To support blind peer-review, or more precisely to protect the novelty of our work until official publication, not all details and files are included yet, but will be.}

\subsection{End-to-End Example}
\label{subsec:illustrative_example}

To demonstrate the full LE flow, we present a compact example using an
original circuit (OC) with three inputs \(a,b,c\) and one primary output:
\[
PO_{OC} = (a \land b) \lor c.
\]

\textbf{Stage \encircle{A}: Construction of the Encrypted Circuit (EC)}

\subsubsection*{A1) Coding Scheme}
We use the NAND/NOR coding scheme:
\[
\text{NAND} \mapsto 0,\qquad \text{NOR} \mapsto 1.
\]

\subsubsection*{A2) Resynthesis and Encoding}
The OC is rewritten using only NAND and NOR gates.  
One valid representation is:
\begin{align*}
t_1 &= \operatorname{NAND}(a,b)', \\
PO_{OC} &= \operatorname{NOR}(t_1,c)'.
\end{align*}

Assume the resynthesized OC contains three gates in the memorized randomized order
\[
[g_1 = \text{NAND},\; g_2 = \text{NAND},\; g_3 = \text{NOR}],
\]
yielding the plaintext
\[
P = 0\,0\,1.
\]

\subsubsection*{A3) AES Encryption of Encoded Gates}
The plaintext is padded to 128 bits and encrypted with AES-128 using a random,
not-stored key:
\[
C = \operatorname{AES}_{K_{\text{AES}}}(P_{\text{padded}}).
\]
For illustration, assume the first three ciphertext bits are:
\[
C = 1\,0\,1.
\]

\subsubsection*{A4) Decoding Ciphertext to Gates}
Each ciphertext bit is decoded using the same codebook:
\[
1 \rightarrow \text{NOR},\qquad
0 \rightarrow \text{NAND}.
\]
Thus the basic EC becomes:
\[
[g_1^{EC}=\text{NOR},\; g_2^{EC}=\text{NAND},\; g_3^{EC}=\text{NOR}].
\]

\subsubsection*{A5) Resynthesis}
The basic EC is resynthesized using the full gate library, producing the
intermediate EC.

\textbf{Stage \encircle{B}: Construction of the Correction Circuit (CC)}

For the single primary output, the CC logic is defined as:
\[
PO_{CC} = PO_{OC} \oplus PO_{EC}',
\]
where \(PO_{EC}'\) is either buffered or inverted.  
Let the randomized choice be inversion, giving:
\[
PO_{EC}' = \overline{PO_{EC}},\qquad K_{CC}=1.
\]
Synthesis yields the final CC.

\textbf{Stage \encircle{C}: Randomization of the EC}

The intermediate EC's primary output is again randomly buffered or inverted.
Assume buffering is selected:
\[
PO_{EC,\text{new}} = PO_{EC},\qquad K_{EC}=0.
\]
Synthesis yields the final EC.

\textbf{Stage \encircle{D}: System-Level Integration}

Each PO of the EC and CC passes through a MUX-based KG structure before entering
the XOR used to produce the final circuit (FC) output:
\[
PO_{FC} = PO_{EC}^{(MUX)} \oplus PO_{CC}^{(MUX)}.
\]

Let the randomized MUX assignments generate:
\[
K_{MUX}(PO_{EC}) = 1,\qquad
K_{MUX}(PO_{CC}) = 0.
\]

\subsubsection*{Final-Key Derivation}
The final key-bits are computed as:
\begin{align}
K_{FC}(PO_{EC}) &= K_{MUX}(PO_{EC}) \oplus K_{EC}
= 1 \oplus 0 = 1, \\
K_{FC}(PO_{CC}) &= K_{MUX}(PO_{CC}) \oplus K_{CC}
= 0 \oplus 1 = 1.
\end{align}

Thus the final key is:
\[
K_{FC} = (1,1).
\]

\textbf{Stage \encircle{E}: Verification}
Formal equivalence checking confirms:
\[
PO_{FC} = PO_{OC},
\]
for the correct final key \(K_{FC}\), completing the end-to-end example.

\section{Experimental Investigation}
\label{sec:experiments}

\subsection{Scope and General Settings}

\begin{table}[tb]
\caption{%
	Basic Properties for Benchmark Circuits
}
\small
\centering
\begin{tabular}{c|c|c|c|c|c}

\multirow{2}{*}{\textbf{Name}}
& \multirow{2}{*}{\textbf{\# Gates}}
& \textbf{Power}
& \textbf{Timing}
& \textbf{Area}
& \textbf{\# Key-Bits}
\\

& %
& \textbf{[$\mu W$]}
& \textbf{[$ns$]}
& \textbf{[$\mu m^2$]}
& \textbf{(This Work)}
\\
\hline
\hline

\textit{c7552}
&726
&711.89
&3.19
&829.65
&214
\\

\rowcolor{Gainsboro}
\textit{c6288}
&709
&1.20e+03
&3.57
&1,287.97
&64
\\

\textit{c5315}
&666
&564.07
&1.12
&740.81
&246
\\

\rowcolor{Gainsboro}
\textit{c3540}
&484
&351.63
&1.52
&521.89
&44
\\

\textit{c2670}
&266
&203.55
&1.05
&297.65
&128
\\

\rowcolor{Gainsboro}
\textit{c1908}
&189
&185.44
&1.15
&210.14
&50
\\

\textit{c1355}
&182
&245.87
&0.76
&234.61
&64
\\

\hline

\rowcolor{Gainsboro}
\textit{b22\_C}
&6,361
&6.74e+03
&5.19
&7,274.30
&1,514
\\

\textit{b21\_C}
&4,274
&4.57e+03
&5.22
&4,759.54
&1,024
\\

\rowcolor{Gainsboro}
\textit{b20\_C}
&4,205
&4.51e+03
&5.10
&4,714.58
&1,024
\\

\textit{b17\_C}
&10,696
&5.46e+03
&5.18
&11,891.26
&2,890
\\

\rowcolor{Gainsboro}
\textit{b15\_C}
&3,295
&1.82e+03
&3.86
&3,666.81
&898
\\

\textit{b14\_C}
&2,061
&1.90e+03
&4.84
&2,254.62
&490
\\

\end{tabular}
\label{tab:benchmarks}

\end{table}

\textbf{Benchmarks:}
All experiments utilize two well-known, fully combinational circuit datasets: ISCAS-85~\cite{hansen1999unveiling} and ITC-99~\cite{davidson1999itc}.
Importantly, this selection enables comparisons with prior art, which most often only support these datasets.
In general, LE is readily compatible with any combinational circuit.
	LE can also support sequential circuits, e.g., when applied between register stages.

All benchmarks are realized in \textit{bench} and \textit{Verilog} format.
For the latter, gate-level netlists are synthesized using the well-known \textit{Nangate 45nm} technology library at typical corner.
Basic properties are listed in Tab.~\ref{tab:benchmarks}. Importantly, the numbers of key-bits are dictated by the workings of LE (Sec.~\ref{sec:method:integration}).

\textbf{Prior Art:}
We compare our novel LE scheme to selected SOTA schemes for LL: TRLL~\cite{trll}, gDMUX~\cite{dmux}, and LUT-L~\cite{baumgarten10}.
We obtain TRLL from the authors~\cite{trll},
gDMUX from~\cite{muxlink_gh}, and LUT-L from~\cite{neos_gh}.
For LUT-L,
    we configure it for LUTs with 2 inputs.
For this and all other schemes, we further utilize their default settings.
For fair comparison, both in terms of security and design overheads, we utilize the same numbers of key-bits (Tab.~\ref{tab:benchmarks}) for all schemes.

\textbf{Dataset Generation I, Randomized Runs:}
Given the inherent randomness for all LL/LE schemes, we conduct multiple runs for each scheme and each benchmark.
For example,
for LE, we conduct 10 independent runs for the encryption stage, and 4 end-to-end runs for each of those, resulting in 40 FC instances per benchmark.
All datasets are used for design analysis and security analysis; details for
the latter are described below.

\textbf{Metrics I, Design Analysis:}
We report average PPA overheads
across all the randomized runs for all schemes.
Area is quantified as standard-cells area, power as total power, and performance/timing as critical-path delay.

\subsection{Settings for Security Analysis}
\label{subsec:security_analysis_setting} 

\textbf{Threat Model:}
Following recent SOTA works, e.g.,~\cite{omla,scope,resynth,muxlink,GNNRE}, we consider the oracle-less threat model. That is, attackers have only access to the protected netlist, but not to an operational IC. Attackers can readily differentiate between regular gates and KG
structures in the protected netlist.

\textbf{Setups for Attacks:}
As indicated, we conduct a thorough security analysis covering major threat vectors for oracle-less attacks.
More specifically, we cover prediction of KG structures by OMLA~\cite{omla}, prediction of KG interconnects by MuxLink~\cite{muxlink}, joint prediction of KG structures and interconnects by resynthesis
with SCOPE~\cite{resynth}, and RE by GNN-RE~\cite{GNNRE};
we utilize respective open-source release for OMLA~\cite{omla_gh}, MuxLink~\cite{muxlink_gh}, resynthesis with SCOPE~\cite{resynth_gh}, and GNN-RE~\cite{gnnre_gh}.\footnote{%
Other attacks like
LIPSTICK~\cite{aghamohammadi2024lipstick} or
NoBALL~\cite{karmakar2024noball} are not released and, thus, not considered here.}

Matching the scope of the various LL/LE schemes against the scope of the various attacks, we apply OMLA for TRLL and LE, MuxLink for gDMUX and LE, and both GNN-RE and resynthesis with SCOPE for all schemes, respectively.
We employ the respective default setups, with customizations as required as follows.

{For GNN-RE, we extend the feature vector to cover all gate types in the full library. We accordingly revise the generation of gate/node labels for circuits.}
For MuxLink on LE,
we assume the attacker can access
the system-level components
individually from within the FC (Fig.~\ref{fig:LE_MUX}).\footnote{%
More specifically,
both the EC and CC are separately available to the attacker.
For each XOR at the final POs,
we model all four possible combinations of interconnects underpinned by the two MUX KG structures related to $K_{MUX}(EC)$ and $K_{MUX}(CC)$.
Moreover, we modify post-processing to consider such link combinations, with the combination having the highest joint probability presumed as the correct one.
Finally, we explore different hop-sizes and find that $h=4$ is most successful; we apply this setting for all MuxLink experiments for fair comparison.}
Accordingly, link prediction is readily accessible to the attacker,
whereas in the real-world, these components would be entangled by the synthesis process for the FC, making such a clear distinction much more difficult.
Thus,
\ul{we conduct a conservative worst-case analysis here.}
For resynthesis with SCOPE on LE, in addition to regular runs on LE where we tackle/attack the FC as synthesized, we follow a similar approach, also inspired by the general divide-and-conquer strategy proposed
in~\cite{resaa}.
As indicated, such worst-case analysis is prudent as the resilience of our LE scheme depends on both the obfuscation of the system-level components (EC and CC) and their interconnects within the FC.

\begin{figure}[tb]
		\includegraphics[width=0.65\columnwidth]{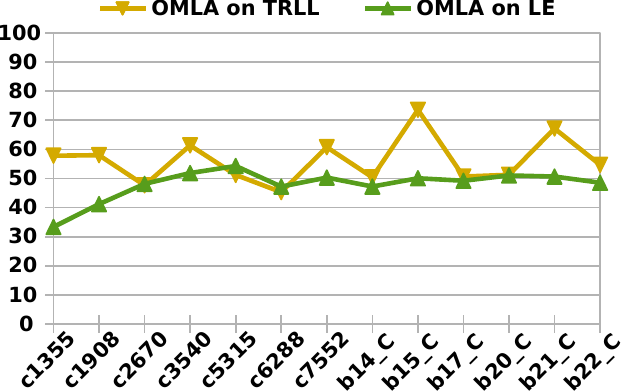}
		\caption{Test accuracy [\%] for OMLA.}
		\label{fig:OMLA}

\end{figure}

\textbf{Dataset Generation II, Security Analysis:}
To enable dataset handling by the various attacks, we follow the respective default settings, with customizations as required as follows.

For GNN-RE, we
obtain a baseline dataset, by resynthesizing each benchmark netlist using various recipes, generating different structural variants for each.
Doing so is important to strive for balancing the different classes/circuits for the library-centric learning approach of GNN-RE.
We follow a similar approach for generation of LL/LE datasets,
i.e., for each scheme, we pick from their randomized runs in a balanced manner, and we also conduct resynthesis (with the same settings, for fairness).
For OMLA, we realize a library-centric training with leave-one-out approach.

\textbf{Metrics II, Security Analysis:}
For the ML-driven attacks OMLA, MuxLink, and GNN-RE, we report the attacks' success rates as average test accuracy.
For GNN-RE, considering the multi-class prediction problem, we also report F1 micro/macro scores.
For baseline/standalone SCOPE as well as resynthesis with SCOPE, we report the attacks' success rates in terms of accuracy (AC) and key-prediction accuracy (KPA).\footnote{%
AC and KPA both quantify the correctly guessed key-bits over all key-bits; the difference is that KPA's denominator excludes unresolved key-bits, whereas AC's includes them.
While AC is the key metric for the attacks' success, KPA provides further insights for the attacks' effectiveness: larger (lower) KPA values indicate more (less) confidence/precision for the resolved key-bits and, vice
versa, more (less) clear decision boundaries against key-bits that cannot be resolved.}
For baseline SCOPE, the key-bit inferences are directly evaluated; for resynthesis with SCOPE, majority voting across all SCOPE runs on all resynthesized circuits is done for each key-bit guess. In both cases, two
complementary sets of key-bit predictions are obtained.
For reporting of both metrics, we first obtain the maxima across these two complementary sets, and then we average across all randomized runs.

\subsection{Security Analysis: KG Structures (OMLA)}
\label{sec:results:sec:OMLA}

The success rate for prediction of the correct KG structures, as in test accuracy for OMLA, is shown in Fig.~\ref{fig:OMLA}.

{LE demonstrates consistently strong resilience across all benchmarks}, with success rates worse than random guessing (i.e., 50\%) for smaller benchmarks and
saturating well around random guessing for larger benchmarks.
{Compared to TRLL, the sole applicable contender for this attack setting, LE is superior by 1.17x on average.}
Furthermore, TRLL is sporadically challenged by larger benchmarks, indicating a lack of resilience scalability; the latter also aligns with more recent findings on TRLL~\cite{insight}.

In short, thanks to its encryption-driven and full-scale obfuscation, LE is more resilient against ML-based prediction of KG structures than prior SOTA (TRLL).

\subsection{Security Analysis: KG Interconnects (MuxLink)}

\begin{figure}[tb]
		\includegraphics[width=0.65\columnwidth]{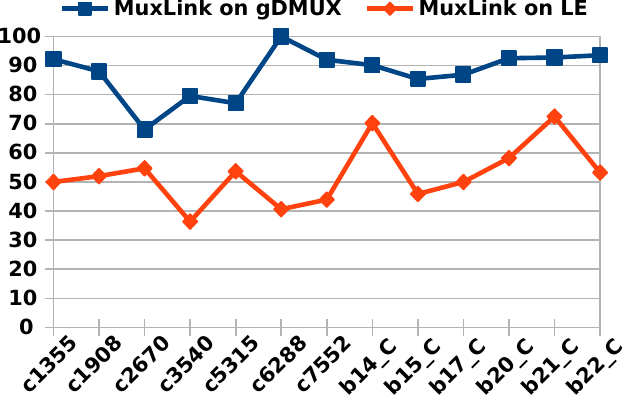}
		\caption{Test accuracy [\%] for MuxLink.}
		\label{fig:ML}

\end{figure}

The success rate for prediction of the correct KG interconnects, as in test accuracy for MuxLink, is shown in Fig.~\ref{fig:ML}.

{LE exhibits good resilience, enforcing an accuracy often close to random guessing.} However, for larger benchmarks, outliers reach to around 70\%, i.e., some correlation exists for the
system-level interconnects between EC and the CC.
This is expected, given that (i)~the EC and CC are orchestrated together to realize the FC and, more importantly, (ii) {we employ a worst-case security analysis here} where those
system-level interconnects
are readily accessible, whereas in the real-world, these are entangled/obfuscated by synthesis of the FC.
Still, {compared to gDMUX, the SOTA for MUX-based LL and sole applicable contender for this attack setting, LE is superior by 1.64x on average.}

In short, thanks to its system-level use of MUX KG structures, LE is more resilient against ML-based prediction of KG interconnects than prior SOTA (gDMUX), even under worst-case settings.

\subsection{Security Analysis: Resynthesis with SCOPE}

\begin{figure}[tb]
		\includegraphics[width=1.03\columnwidth]{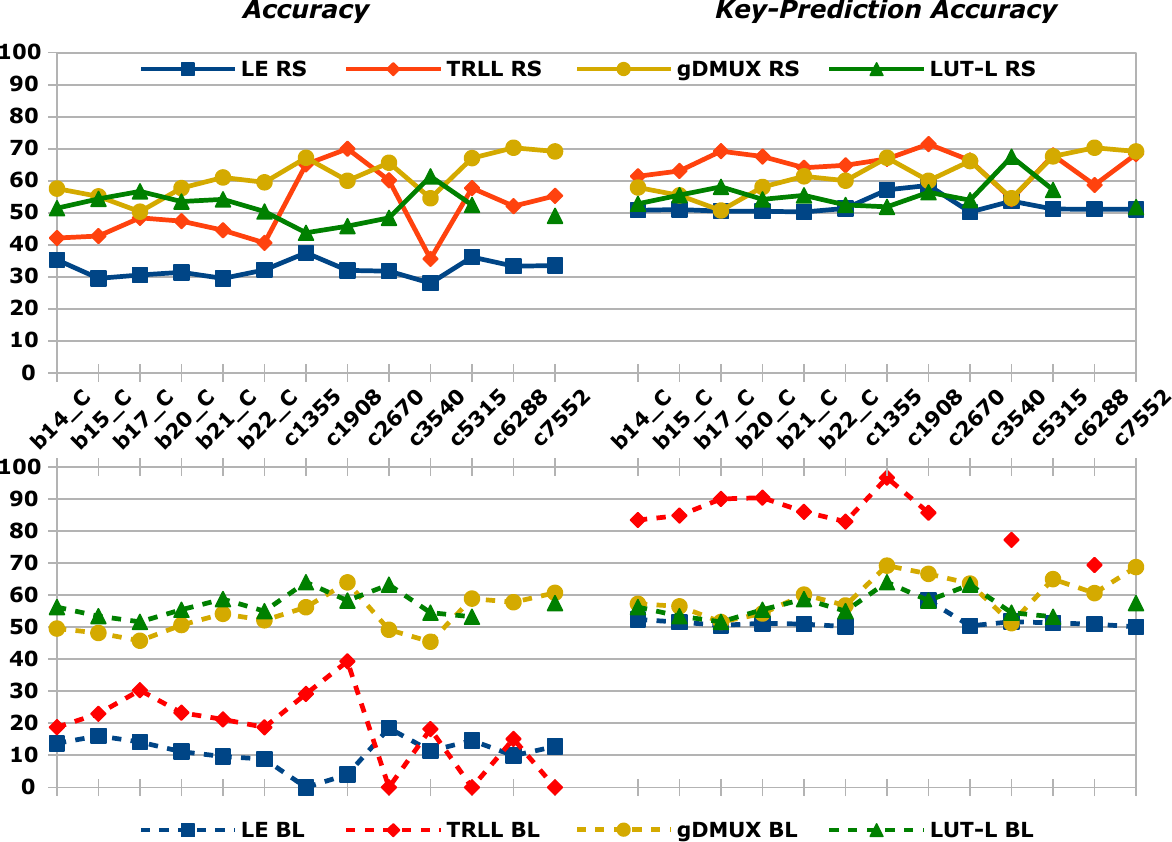} 
		\caption{Accuracy and key-prediction accuracy [\%] for resynthesis with SCOPE (RS; top) and baseline SCOPE (BL; bottom), respectively.
For KPA, data points are missing for cases with AC=0\%, where KPA is undefined.
For LUT-L, for the benchmark c6288, data points are missing as the correct key
was not returned by the related tool~\cite{neos_gh}.}
		\label{fig:resynth}

\end{figure}

The success rates for
joint prediction of KG structures and interconnects by
baseline/standalone SCOPE and resynthesis with SCOPE, respectively, are shown in Fig.~\ref{fig:resynth}.

For the most relevant results, namely for AC for resynthesis with SCOPE, {LE outperforms, on average, TRLL by 1.57x, gDMUX by 1.89x, and LUT-L by 1.60x, respectively.}
For AC for baseline SCOPE, LE even outperforms TRLL by 1.64x, gDMUX by 4.79x, and LUT-L by 5.10x, respectively, which
implies that both gDMUX and LUT-L are already challenged by baseline SCOPE attacks, whereas LE (and LUT-L) induce more complex obfuscation, justifying further attack efforts via resynthesis with SCOPE.
Importantly, {LE is the only scheme that forces these further efforts well below the random-guessing threshold across all benchmarks}, i.e., its resilience scales well
for this sophisticated attack setting.
For KPA for resynthesis with SCOPE vs.\ baseline SCOPE attacks, LE outperforms TRLL by 1.25x vs.\ 1.64x, gDMUX by 1.18x vs.\ 1.16x, and LUT-L by 1.07x vs.\ 1.10x, respectively, all on average.
Again, LE is the only scheme that pushes KPA toward random-guessing domains across all benchmarks.
This reconfirms the resilience of LE, as in resynthesis efforts are not contributing much toward more clear decision-making for SCOPE when resolving the key-bits.

Upon further investigation, we note the following for LE.
The construction of both EC and CC components, and their system-level integration with MUX KG structures, does
not provide much leverage for synthesis-related attack efforts.
This is because the integration of both the EC and CC is essentially based on whether POs/cones are used as-is vs.\ inverted at the system level, which is different from
embedding KG structures throughout the netlist, as done by prior art in LL.
Accordingly, for LE,
the underlying circuit structures do not change significantly under SCOPE ``hard-coding'' the different possible combinations of key-bit values.

\begin{figure}[tb]
		\includegraphics[width=1.03\columnwidth]{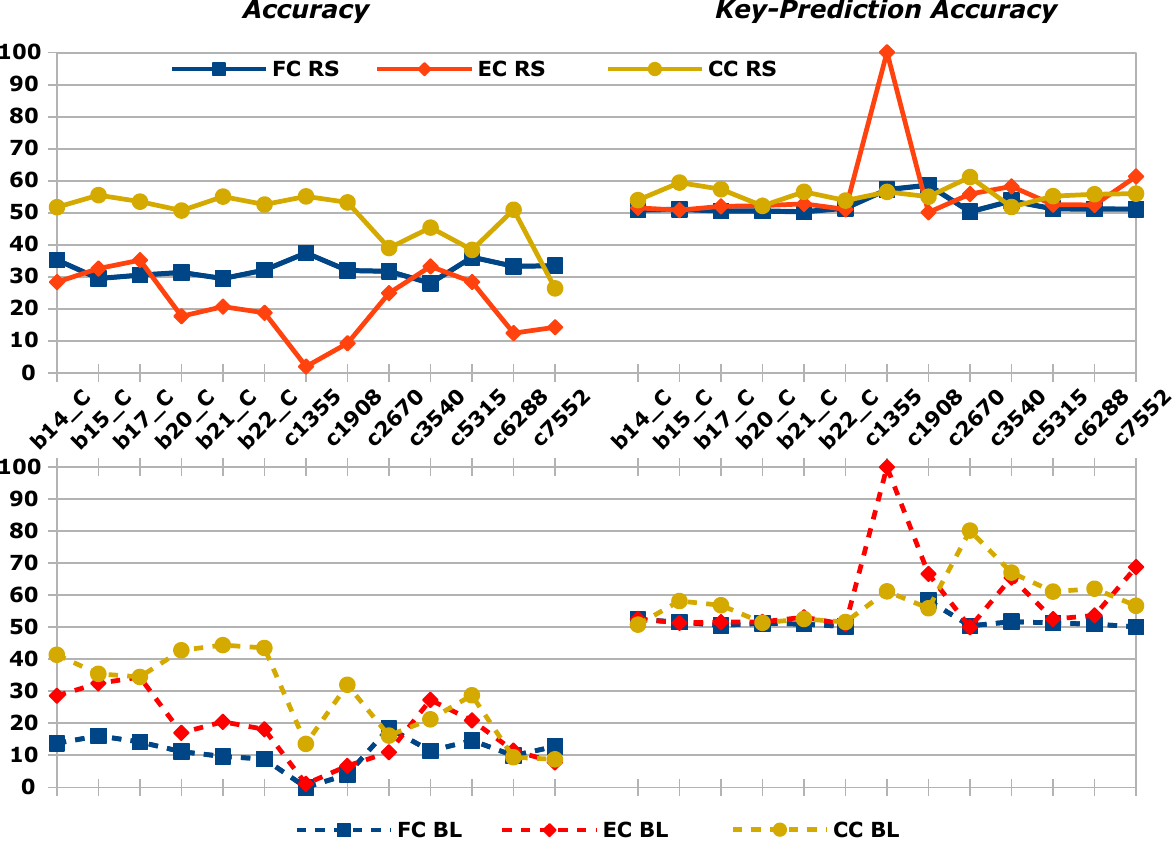} 
		\caption{Accuracy and key-prediction accuracy [\%] for attacking LE via resynthesis with SCOPE (RS; top) and baseline SCOPE (BL; bottom), respectively, all for the worst-case settings of attacking the EC
			and CC separately.
				Note that FC results are for the regular setting of attacking LE as is, carried over from Fig.~\ref{fig:resynth}.
For KPA, data points are missing for cases with AC=0\%, where KPA is undefined.}
		\label{fig:resynth-LE-worst}

\end{figure}

In Fig.~\ref{fig:resynth-LE-worst}, we provide further results for the outlined worst-case setting, i.e.,
without the system-level obfuscation realized by synthesizing the FC,
where both EC and CC are attacked directly and separately for their respective intermediate key-bits,
following the general divide-and-conquer strategy proposed in~\cite{resaa}.

For the most relevant results, i.e., AC for resynthesis with SCOPE, attacking the EC achieves 21.44\% AC (i.e., only 0.66x of the AC achieved when attacking the FC, or ``AC for FC'' for short), whereas
attacking the CC achieves 48.35\% AC (i.e., 1.49x AC for FC), all on average.
For AC for baseline SCOPE, attacking the EC achieves 18.22\% (i.e., 1.64x AC for FC), whereas attacking the CC achieves 28.58\% (i.e., 2.57x AC for FC), respectively.
These results are expected; there should be some benefit for attackers tackling the EC and CC directly and separately.
Still, for KPA, values are close to random guessing, aside from the outlier for benchmark c1355, where the 100\% KPA is a skewed result that arises from very few key-bits (i.e., 1 to 2) resolved at all, which shows from the
corresponding AC value close to zero.
As before, resynthesis efforts are not contributing much toward more clear decision-making for SCOPE.

In addition to these insights when compared to attacking the FC, these results also show that (i)~the EC is more resilient than the CC, (ii)~resynthesis efforts do not scale well for attacking the EC, and (iii), most importantly, results remain in the random-guessing domain.
Upon further investigation, we note the following.
First, recall that results relate to the respective intermediate key-bits; thus, while the CC is more vulnerable when isolated, security analysis still requires a holistic view.
That is, recovering the key for one component is insufficient; an attacker must successfully break both the EC and the CC.
Second, circuit structures of the EC tend to be smaller and less complex than those of the CC, which can be explained by their respective construction steps:
the EC is fully obfuscated by encryption, providing ample opportunities for
synthesis to simplify the related circuitry, whereas the CC, while also derived from the EC, needs to realize restoration of the OC's functionality, which tends to become more complex.
These aspects can explain the stronger resilience of the EC.

In short, thanks to the design and construction of both the EC and CC and their system-level integration with MUX KG structures, LE is more resilient against joint prediction of KG structures and interconnects than prior
SOTA of LL.
This holds true even for the
worst-case setting where attackers can directly and separately tackle the EC and CC.
This resilience is because the circuit structures of both EC and CC are operated either as-is vs.\ inverted at the system level, thereby remaining
largely unaffected by synthesis-related restructuring efforts, which is different from the construction and operation of LL schemes.

\subsection{Security Analysis: Reverse Engineering (GNN-RE)}
\label{subsec:security_analysis_gnnre}

\begin{figure}[tb]
\centering
		\includegraphics[width=1.0\columnwidth]{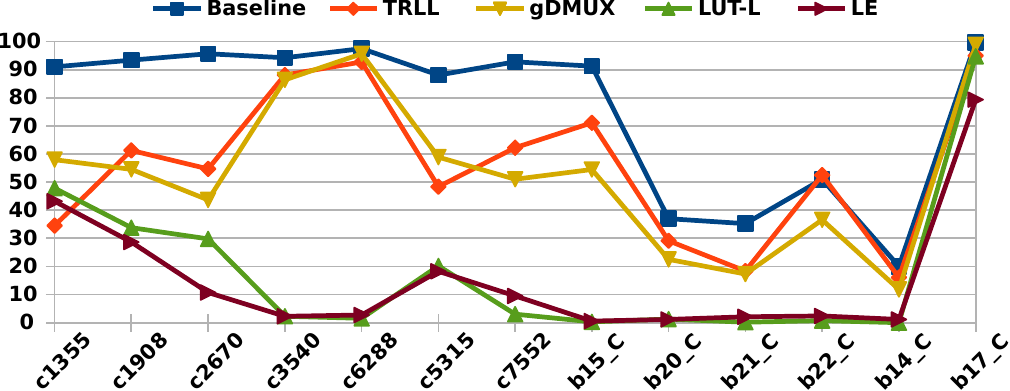} 
		\caption{GNN-RE test accuracy [\%].}
		\label{fig:GNN-RE}

\end{figure}

The success rate for reverse engineering, as in
test accuracy for GNN-RE, is shown in Fig.~\ref{fig:GNN-RE}.
We also report on the related multi-class prediction performance in Tab.~\ref{tab:GNNRE_metrics}.
Note that baseline refers to the benchmarks as-is, i.e., without any LL/LE applied for obfuscation.

\begin{table}[tb]
\caption{%
	GNN-RE Multi-Class Prediction Metrics
    }
\small
\centering
\setlength{\tabcolsep}{2.0pt} 
\begin{tabular}{c|c|c|c|c|c}

& \textbf{Baseline}
& \textbf{TRLL~\cite{trll}}
& \textbf{gDMUX~\cite{dmux}}
& \textbf{LUT-L~\cite{baumgarten10}}
& \textbf{LE [This]}
\\
\hline
\hline

\textit{F1 Micro}
&72.17\%
&60.16\%
&56.60\%
&28.81\%
&25.66\%
\\

\rowcolor{Gainsboro}
\textit{F1 Macro}
&75.47\%
&55.85\%
&53.96\%
&14.40\%
&11.60\%
\\

\end{tabular}
\label{tab:GNNRE_metrics}

\end{table}

LE demonstrates a markedly stronger ability to obstruct GNN-RE,
imposing the lowest
F1 micro and macro scores, i.e., the lowest prediction performance, among all schemes (Tab.~\ref{tab:GNNRE_metrics}).
Averaging the test accuracy across all benchmarks in Fig~\ref{fig:GNN-RE}, {LE outperforms TRLL by 3.58x, gDMUX by 3.41x, and LUT-L by 1.17x, respectively}.

We also observe that {LE (and LUT-L) exhibit strong resilience
for both larger benchmarks with larger key-sizes and smaller ones.}
In contrast, TRLL and gDMUX face significant challenges
when dealing with larger instances.
The latter is because TRLL and gDMUX primarily introduce structural changes localized around the inserted KG structures, limiting their ability to
obscure the entire design.
Finally, note that b17\_C is much more easily identified by GNN-RE across the board. This is due to the underlying functional similarities across ITC benchmarks~\cite{davidson1999itc}.
LE still outperforms prior art for this challenging benchmark, namely by 1.21x on average.

In short,
thanks to its encryption-driven and full-scale obfuscation, LE is more resilient against ML-based RE than prior SOTA.

\subsection{Security Analysis: Discussion on Oracle-Guided Attacks}
\label{sec:disc}

While conventional wisdom suggests oracle-guided attacks are stronger -- due to the attacker's additional capabilities through oracle access -- we may argue that the reverse holds when evaluating defenses.
That is,
	     some LL/LE scheme is powerful
when readily resisting oracle-less attacks, whereas potential resilience against oracle-guided attacks only represents a weaker defense posture, as the latter assumes greater attacker capabilities.
In other words,
	security assessment from the defender's viewpoint may rate attack models requiring less (vs more) resources or capabilities as more (vs less) powerful.
As we propose a defense, we adopt this perspective on
oracle-less attacks for our security analysis.

Nevertheless, in future work, we will investigate LE also in the oracle-guided threat model.\footnote{%
	Since doing so would require us to obtain, configure, run, and evaluate our experiments for various other LL schemes proposed against oracle-guided attacks, such efforts
would be considerable and, again, considered as scope for future work.}
The fact that LE incorporates entanglement of system-level interconnects is potentially promising toward construction of SAT-hard
problem instances, as outlined in~\cite{interlock}.
Besides, LL/LE schemes can be extended
via compound locking against
oracle-guided attacks~\cite{tehranipoor2024advances,chakraborty2019keynote}, although doing so still requires some careful security analysis~\cite{tan2020benchmarkingfrontierhardwaresecurity,9474118,resaa}.

\subsection{Design Analysis}

\begin{figure*}[tb]
\centering
		\includegraphics[width=0.85\textwidth]{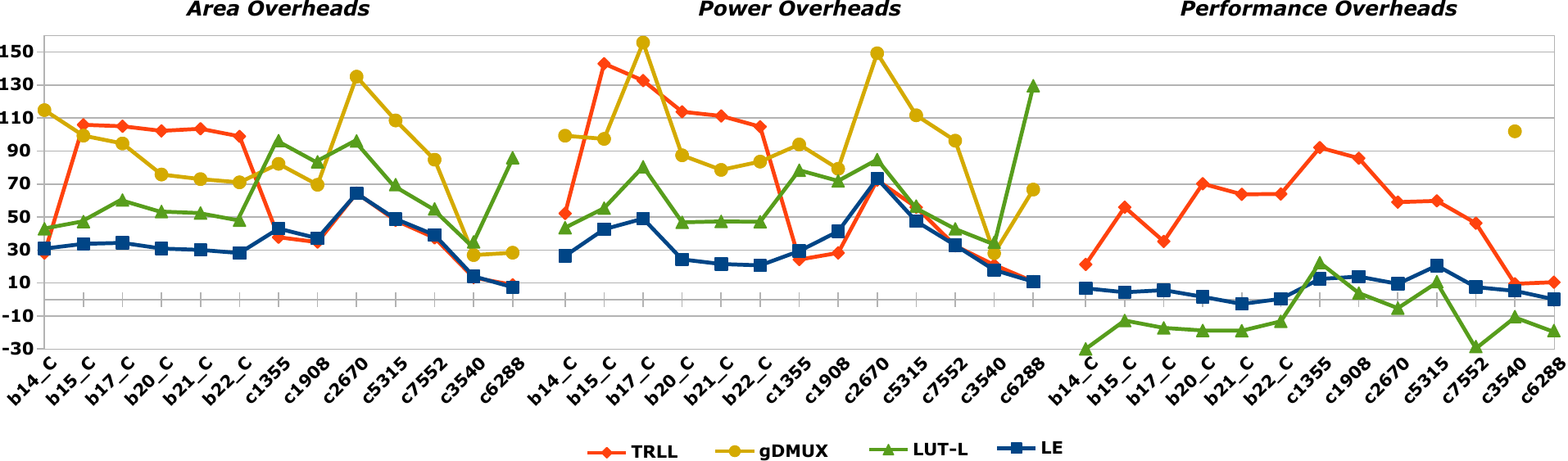} 
		\caption{Average PPA overheads [\%] for all considered schemes.
				Note that datapoints not shown (i.e., for performance for gDMUX) are exceeding the upper limit of the plot, i.e., 160\%.}
		\label{fig:PPA}
\end{figure*}

PPA overheads are shown in Fig.~\ref{fig:PPA}.
Our proposed LE scheme significantly outperforms prior art for both area and power, and is a competitive runner-up against LUT-L for performance.
More specifically, across the various benchmarks,
area overheads over LE are
1.78x for TRLL, 2.41x for gDMUX, and 1.85x for LUT-L, respectively;
power overheads are
2.06x for TRLL, 2.80x for gDMUX, and 1.87x for LUT-L, respectively;
and performance overheads are
7.86x for TRLL, 71.56x for gDMUX,
and -1.63x for LUT-L, respectively.
{When looking at the ``bigger picture'' of combined PPA overheads} (i.e., overheads for area, power, and performance are summed up separately for each benchmark), 
{LE outperforms, on average,
   TRLL by 2.45x, gDMUX by 8.72x, and LUT-L by 1.55x, respectively.}

We observe that {PPA overheads for LE}, and to some degree also for LUT-L, {often scale graciously for both larger circuits and key-sizes as well as smaller ones}. In contrast, for power and area, TRLL is challenged by larger instances and gDMUX
is sporadically challenged by both larger and smaller instances.
For performance, TRLL is more challenged by smaller instances, whereas gDMUX shows excessive overheads, namely 471.63\% on average.
The latter is because gDMUX selects data inputs for its MUX-based KG structures
throughout the entire netlist. Thus, especially for larger key-bit sizes, most timing paths become iteratively entangled by gDMUX, which significantly lengthens
the critical path.
In contrast, critical paths are not impacted much for LE,
as the main components (EC and CC) are orchestrated in parallel.

In short, thanks to its simple circuit design and synthesis-driven implementation -- and despite its full-scale obfuscation -- LE is competitive against SOTA in terms of PPA overheads.

\section{Discussion}
\label{sec:discussion}

Roy \emph{et al.}'s EPIC framework~\cite{roy2008epic} was the first to incorporate cryptography into logic protection, but its use is restricted to the secure \emph{delivery} of the activation key. EPIC embeds key-programmable gates into the design and uses asymmetric cryptography (e.g., RSA) only to encrypt the key that enables correct operation; once the key is loaded, the circuit structure is unchanged and fully visible, as no part of the logic itself is encrypted or structurally transformed. Later works apply cryptography-inspired mechanisms at finer granularity. \emph{Encrypt Flip-Flop}~\cite{karmakar2018encrypt} protects sequential logic by inserting MUXes at selected flip-flop outputs; key-bits select between true and complemented values so that incorrect keys propagate corrupt state values, yet the combinational logic structure remains entirely exposed. \emph{Hardware Functional Obfuscation}~\cite{yu2022hardware} introduces FeFET-based active interconnects that can be programmed to behave as an inverter or buffer, allowing small logic regions to be reconfigured at run time. Although this provides a hardware-level means of concealing local behavior unless the correct configuration is programmed, only limited logic blocks are affected, and the original netlist is otherwise left intact. Across these works, cryptography or logic programmability is used to guard specific key-controlled elements, not to encrypt or transform the full logic representation.

Our LE methodology employs cryptography to protect the \emph{entire} logic rather than isolated key-gates or reconfigurable segments. We encode the full gate-level description of the design, apply a block cipher (e.g., AES) to produce ciphertext that directly determines the encrypted circuit structure, and decode this ciphertext into a synthesized encrypted circuit whose internal gate types, connectivity, and functional behavior are all cryptographically randomized. This process obfuscates the circuit end-to-end: the logic is rearranged, structurally altered, and rendered unintelligible without the decryption machinery and correct key. Whereas prior works preserve the original netlist and merely alter its activation or local behavior, our approach remodels the netlist itself under cryptographic control, achieving comprehensive logic-level obfuscation capable of resisting stronger adversaries, including oracle-less reverse engineering, structural analysis, and resynthesis-based attacks.

\section{Conclusions and Future Work}

We have presented a first-of-its-kind work for actual LE.
Recall that LL handles the integration of KG
structures
more as an afterthought, leaving them vulnerable to various attacks.
In contrast, our approach to LE
truly obfuscates the circuit's structure and functionality at full scale.
With our extensive analysis, we demonstrate the promise of our LE scheme.
Ours consistently outperforms prior art against a range of SOTA oracle-less attacks covering crucial threat vectors, all with lower PPA overheads.
For example, LUT-L~\cite{baumgarten10}, the only contender to some degree, is still inferior to LE, namely by 1.17x for GNN-RE, by 1.60x/5.10x for resynthesis and/or SCOPE, and by 1.55x for PPA, respectively.
The fact that LUT-L utilizes redaction and by doing so differs from traditional LL approaches reiterates the fallacies of prior art.
LE excels over LUT-L thanks to its system-level obfuscation, whereas LUT-L is still limited to local redaction/obfuscation.
Finally, we provide a full open-source release \cite{release}, fostering reproducibility and future work.

      We envision multiple directions for future work with LE.
First, we shall explore to also encode and encrypt the interconnects, together with the logic/gates.
Second, we shall extend LE for the oracle-guided threat model.
Third, we shall apply LE for larger and sequential circuits,
including modern
chiplet systems.\footnote{%
	Our LE method lends itself readily to the latter:
the EC and the CC can be realized as two third-party chiplets, whereas secure system
integration can be subsequently and separately realized by an active interposer~\cite{park20}.
Notably, such an implementation and supply-chain setup would leave the chiplet providers, which may be acting malicious, without access to the system-level oracle, thereby still hindering oracle-guided attacks.}

\bibliographystyle{unsrt}
\bibliography{DAC/reference.bib}

\end{document}